\documentclass[showpacs,prl,aps,twocolumn,floatfix]{revtex4}

\usepackage{bm}
\usepackage{amsmath}
\usepackage{amssymb}
\usepackage{latexsym} 
\usepackage{amsfonts} 
\usepackage{epsfig}
\usepackage{psfrag} 

\newcommand{\ket}[1]{\left|#1\right>} 
\newcommand{\bra}[1]{\left<#1\right|}

\newcommand{\f}[1]{\mbox{\boldmath$#1$}}
\newcommand{\fk}[1]{\mbox{\boldmath$\scriptstyle#1$}}

\newcommand{\na}{\mbox{\boldmath$\nabla$}}
\newcommand{\bea}{\begin{eqnarray}}
\newcommand{\ea}{\end{eqnarray}}
\newcommand{\eea}{\end{eqnarray}}
\newcommand{\ord}{{\cal O}}

\begin{document}
  
\title{Table-top creation of entangled multi-keV photon pairs via the 
  Unruh effect} 

\author{Ralf Sch\"utzhold$^{1,*}$, Gernot Schaller$^1$, 
and Dietrich Habs$^2$}    

\affiliation{$^1$Institut f\"ur Theoretische Physik, 
Technische Universit\"at Dresden, 01062 Dresden, Germany}  

\affiliation{
$^2$Department f\"ur Physik der Ludwig-Maximilians-Universit\"at M\"unchen
und Maier-Leibnitz-Laboratorium, 
Am Coulombwall 1, 85748 Garching, Germany
}

\begin{abstract}
Electrons moving in a strong periodic electromagnetic field 
(e.g., laser or undulator) may convert quantum vacuum fluctuations 
into pairs of entangled photons, which can be understood as a
signature of the Unruh effect. 
Apart from verifying this striking phenomenon, the considered effect
may allow the construction of a table-top source for entangled
photons (``photon pair laser'') and the associated quantum-optics
applications in the multi-keV regime with near-future facilities. 
\end{abstract}


\maketitle

The striking discovery that the particle concept in quantum field
theory may depend on the inertial state of the observer is one of the
main lessons from the Unruh effect:
The Minkowski vacuum is the ground state 
with respect to all stationary and inertial observers 
(moving with a constant velocity).
However, an accelerated (i.e., non-inertial) observer generally
experiences the Minkowski vacuum as an excited quantum state with a
non-vanishing particle content.
In case of uniform acceleration $a$, it corresponds to a thermal
bath characterized by the Unruh temperature \cite{unruh}
\bea
\label{unruh}
T_{\rm Unruh}
=
\frac{\hbar}{2\pi k_{\rm B}c}\,a
\,.
\ea
Now, considering an accelerated electron, for example, there is a 
finite probability that a comoving (non-inertial) observer witnesses 
the scattering of a  photon out of the thermal bath by the electron 
(due to its nonzero Thomson cross section).  
Translation of this scattering event in the accelerated frame into the 
(inertial) laboratory frame corresponds to the emission of pair of
real photons \cite{happen}.
Therefore, accelerated electrons may convert (virtual) quantum 
vacuum fluctuations into real particle pairs \cite{chen} via
non-inertial scattering -- which can be understood as a signature of
the Unruh effect (similar to moving-mirror radiation \cite{mirror}).   
In a previous work \cite{habs}, we studied electrons under the influence
of an approximately constant electric field (corresponding to the case
of uniform acceleration) and found that these signatures might be 
detectable for field strengths not too far below the Schwinger limit
\cite{schwinger}.  

In the following, we shall focus on an alternative set-up 
(nonuniform acceleration) and consider electrons which are shot with
ultra-relativistic velocities into a strong periodic (e.g., harmonic) 
electromagnetic field, such as a laser beam or an undulator.
In the rest frame of the ultra-relativistic electrons, the
(transversal) field strength is strongly boosted and thus the
acceleration felt by the electrons is amplified.
During each acceleration cycle, the electrons emit a small amplitude
for photon pair creation and all these amplitudes may add up
constructively. 

In order to demonstrate the main idea, let us assume that the frequency 
$\omega$ (measured in the rest frame of the electrons) 
of the external electromagnetic field $E,B$ lies far below the electrons 
rest mass $m\gg\omega$ and that its normalized amplitude is much smaller 
than one 
\bea
\label{normalized}
q^2E^2+q^2B^2\ll m^2\omega^2 
%
\,,
\ea
where $q$ is charge of electron. 
In the natural units $\hbar=c=\varepsilon_0=\mu_0=1$ used here, it is 
related to the fine-structure constant $\alpha_{\rm QED}$ via 
$q=\sqrt{4\pi\alpha_{\rm QED}}\approx0.3$.  
In the rest frame of the electrons, their classical quivering motion
induced by the external field  
\bea
\label{quivering}
\f{r}_{\rm cl}(t)=\f{e}_z\,\frac{qE}{m\omega^2}\,\cos(\omega t)
\ea
is nonrelativistic $\f{\dot r}_{\rm cl}^2\ll1$ and thus the impact 
of the magnetic field (i.e., photon pressure) can be neglected.
Furthermore, the spin of the electrons can be ignored since the 
spin energy $\mu_eB$ is much smaller than the frequency 
$\mu_eB\ll\omega$. 
Hence the dynamics of the electrons under the influence of the
(classical plus quantum) electromagnetic field is governed by the 
Lagrangian 
\bea
L(\f{\dot r}_e,\f{r}_e)
=
\frac{m}{2}\,\f{\dot r}^2_e-q\f{\dot r}_e\cdot\f{A}(\f{r}_e)
\,,
\ea
where $\f{A}$ is the vector potential in temporal gauge.  
Now we split the electromagnetic field 
$\f{A}=\f{A}_{\rm cl}+\f{A}_{\rm qu}$ into a large classical
part $\f{A}_{\rm cl}$ plus small quantum fluctuations 
$\f{A}_{\rm qu}$, e.g., scattered photons.   
Accordingly, the electron trajectory 
$\f{r}_e=\f{r}_{\rm cl}+\f{r}_{\rm qu}$ is split up into the 
classical quivering motion $\f{r}_{\rm cl}$ in Eq.~(\ref{quivering}) 
plus small quantum fluctuations $\f{r}_{\rm qu}$ due to coupling to the 
quantized electromagnetic field $\f{A}_{\rm qu}$.
From the Euler-Lagrange equations
\bea
\frac{d}{dt}
\left[
m\f{\dot r}_e-q\f{A}(\f{r}_e)
\right]
=-
q\frac{\partial}{\partial\f{r}_e}
\left[
\f{\dot r}_e\cdot\f{A}(\f{r}_e)
\right]
\ea
we see that the canonical momentum $\f{p}_e=m\f{\dot r}_e-q\f{A}$ is 
conserved to first order $\f{p}_{\rm qu}$ if the right-hand
side vanishes. 
This is precisely the condition for planar Thomson scattering which 
is satisfied if the polarizations are orthogonal or, alternatively, 
for planar momenta $\f{k}_{\rm qu},\f{k}_{\rm cl}$
\bea
\f{A}_{\rm qu}
\perp 
\f{A}_{\rm cl}
\,\|\,
\f{r}_{\rm cl}
\perp 
\f{r}_{\rm qu}
\,\vee\,
\f{k}_{\rm qu}
\perp 
\f{r}_{\rm qu}
\,\|\,
\f{A}_{\rm qu}
\perp 
\f{k}_{\rm cl}
\,.
\ea
In this case, we get (up to an irrelevant constant)
\bea
\label{momentum}
\f{\dot r}_{\rm qu}=\frac{q}{m}\,\f{A}_{\rm qu}
\,.
\ea
Now let us consider the equations of the electromagnetic 
field depending on the full electron trajectory $\f{r}_e(t)$
\bea
\label{source}
\f{\ddot A}-\na\times(\na\times\f{A})
=
-q\f{\dot r}_e\delta^3(\f{r}_e-\f{r})
\,.
\ea
(The longitudinal component $\na\cdot\f{A}$ is non-vanishing in
temporal gauge and contains the instantaneous Coulomb field, which
does not contribute to the radiation content.)
Inserting the split $\f{r}_e=\f{r}_{\rm cl}+\f{r}_{\rm qu}$ 
into the source term yields the classical Larmor radiation from 
$\f{r}_{\rm cl}$ plus quantum corrections.
There are two lowest-order corrections: 
variations of the electron position 
$\delta^3(\f{r}_{\rm cl}+\f{r}_{\rm qu}-\f{r})$
plus the current $\f{\dot r}_{\rm qu}$ due to quantum fluctuations. 
Since the first contribution vanishes for parallel photons $\f{k}\|\f{k'}$ 
(the case we are mostly interested in) and does not generate polarization
correlations (which will be used for detection), we shall focus on the
quantum current $\f{\dot r}_{\rm qu}$.  
Combining Eqs.~(\ref{momentum}) and (\ref{source}), we obtain the effective 
interaction Hamiltonian for planar Thomson scattering 
\bea
\hat H_{\rm eff}(t)=\frac{q^2}{2m}\,\f{\hat A}^2[t,\f{r}_{\rm cl}(t)]
\,.
\ea
(Note that a factor of 2 is missing in \cite{habs}.)
The photon pairs created out of the quantum vacuum by non-inertial 
scattering can now be calculated via time-dependent perturbation theory 
yielding the two-photon amplitude 
\bea
\label{two-photon-full}
{\mathfrak A}_{\fk{k},\lambda,\fk{k'},\lambda'}
=
\frac{q^2}{4m}\,
\frac{\f{e}_{\fk{k},\lambda}\cdot\f{e}_{\fk{k'},\lambda'}}{V\sqrt{kk'}}\,
{\mathcal F}_{\fk{k},\fk{k'}}
\,,
\ea
where $V$ is the quantization volume and $\f{e}_{\fk{k},\lambda}$,  
$\f{e}_{\fk{k'},\lambda'}$ denote the (linear) polarization vectors of
the two created photons with wave-numbers $\f{k}$ and $\f{k'}$,
respectively.  
The remaining time integral,  
\bea
\label{time-int}
{\mathcal F}_{\fk{k},\fk{k'}}
=
i\int dt\,
\exp\left\{i(k+k')t-i(\f{k}+\f{k'})\cdot\f{r}_{\rm cl}(t)\right\}
\,,
\ea
can be Taylor expanded for small oscillation amplitudes 
\bea
{\mathcal F}_{\fk{k},\fk{k'}}
\approx
\int dt\,
e^{i(k+k')t}
(\f{k}+\f{k'})\cdot\f{r}_{\rm cl}(t)
\,,
\ea
and just yields the Fourier transform of the quivering motion
$(\f{k}+\f{k'})\cdot\f{\tilde r}_{\rm cl}(k+k')$ 
evaluated at a frequency of $k+k'$ and projected onto 
$\f{k}+\f{k'}$.
The resonance condition (energy conservation) reads $k+k'=\omega$ and 
at resonance $k+k'=\omega$, the amplitude yields 
\bea
\label{two-photon}
{\mathfrak A}_{\fk{k},\lambda,\fk{k'},\lambda'}
=
\frac{q^3E}{8m^2}\,
\frac{\f{e}_{\fk{k},\lambda}\cdot\f{e}_{\fk{k'},\lambda'}}{\omega^3V}\,
\frac{k_z+k_z'}{\sqrt{kk'}}\,
\omega T
\,,
\ea
where $\omega T$ counts the number of laser cycles experienced by the 
electrons. 
The probability of emitting a pair of photons in resonance band
$k+k'=\omega\pm\ord(1/T)$ can be estimated via 
$\sum_{\lambda,\lambda'}
(\f{e}_{\fk{k},\lambda}\cdot\f{e}_{\fk{k'},\lambda'})^2=
1+(\f{e}_{\fk{k}}\cdot\f{e}_{\fk{k'}})^2\geq1$ 
\bea
\label{prob-unruh}
{\mathfrak P}_{\rm Unruh}
>
\frac{\alpha_{\rm QED}^2}{(4\pi)^2}
\left[\frac{E}{E_S}\right]^2
\times
\ord\left(
\frac{\omega T}{30}
\right)
\ll1
\,,
\ea
where $E_S=m^2/q$ denotes the Schwinger limit \cite{schwinger} and the
exact pre-factor depends on the pulse shape etc.
\footnote{Note that, after reaching the relativistic limit $qE=\omega
  m$, the probability ${\mathfrak P}_{\rm Unruh}$ scales with
  $\omega^2/m^2$,  
  i.e., further increasing $E$ basically does not enhance the
  probability ${\mathfrak P}_{\rm Unruh}$ for the lowest resonance
  anymore, merely the higher harmonics grow.}.

Of course, the electron does not just act as a scatterer, but also 
possesses a charge -- and, as every accelerated charge, emits Larmor 
radiation.
This classical radiation corresponds to a coherent state and can fully
be described by the associated one-photon amplitude 
\bea
\label{Larmor}
\alpha_{\fk{k},\lambda}
=
q\int dt\,
\frac{\f{e}_{\fk{k},\lambda}\cdot\f{\dot r}_{\rm cl}(t)}{\sqrt{2Vk}}\,
\exp\{ikt-i\f{k}\cdot\f{r}_{\rm cl}(t)\}
\,.
\ea
From the scalar product 
$\f{e}_{\fk{k},\lambda}\cdot\f{\dot r}_{\rm cl}$, one may read off the
well-known blind spot and the fixed polarization. 
Similarly to the above estimate (\ref{prob-unruh}), the one-photon
probability of this classical counterpart yields  
\bea
\label{prob-Larmor}
{\mathfrak P}_{\rm Larmor}^{(1)} 
=
\frac{\alpha_{\rm QED}}{2\pi}
\left[\frac{qE}{m\omega}\right]^2
\times
\ord\left(
\frac{\omega T}{3}
\right)
\,.
\ea
In view of $\omega\ll m$, the total classical probability above 
exceeds the probability ${\mathfrak P}_{\rm Unruh}$ of quantum
radiation. 
However, as one may infer from Eq.~(\ref{Larmor}), the classical 
resonance condition reads $k=\omega$, i.e., the Larmor photons are 
predominantly monochromatic (in the electron frame).
In contrast, the photon pairs created via the Unruh effect occur at 
different frequencies, as long as they satisfy $k+k'=\omega$, i.e., 
these pairs are correlated in energy and polarization.  
(Whereas Larmor radiation has a fixed polarization and a blind spot 
in $z$-direction.)  
Note that the ratio of the probabilities in Eqs.~(\ref{prob-unruh})
and (\ref{prob-Larmor}) is roughly independent of the field strength
$E$ given by 
\bea
\label{ratio}
\frac{{\mathfrak P}_{\rm Unruh}}{{\mathfrak P}_{\rm Larmor}^{(1)}}
=
\ord\left(
\frac{\alpha_{\rm QED}}{80\pi}\,
\frac{\omega^2}{m^2}
\right)
\,.
\ea
%

\begin{figure}[ht]
\includegraphics[height=4.2cm]{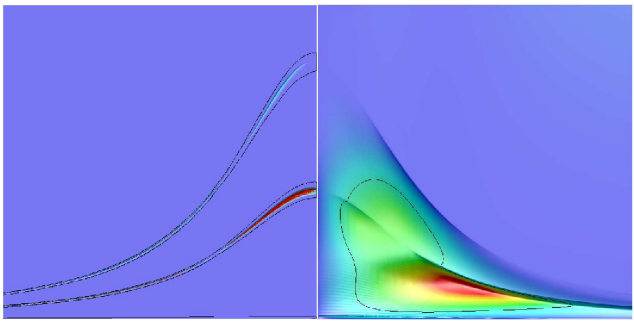}
\caption{\label{figure1} $E$-$\vartheta$ plot of the one-photon
  probability of classical (Larmor, left half of image) and quantum
  (Unruh, right) radiation in the laboratory frame. An electron with a
  boost factor of $\gamma=300$ hits a counter-propagating optical
  Gaussian laser pulse with an intensity around $10^{18}\rm W/cm^2$ and
  a half-width of 100 cycles. The photon energy $E$ ranges from zero
  (bottom) to 2 MeV (top) and $\vartheta$ varies from zero (middle) to
  1/100 (left and right boundary). In the chosen color coding (not the
  same in the two images), red indicates a large and dark blue a vanishing
  probability. The black iso-lines denote the same values in both
  pictures and show that the quantum radiation dominates in certain
  phase space regions (which could be extracted with apertures and
  energy filters). For example, cutting out a small cone around the
  blind spot (at $\vartheta=1/300$ in the left picture) drastically
  increases the Unruh-Larmor ratio (\ref{ratio}).  
  As one may infer from the visibility of the
  second harmonic, relativistic effects already start to play a role
  for this set of parameters ($\f{\dot r}_{\rm
  cl}^2\approx1/9$). Therefore, we numerically calculated the full
  electron trajectory (including the impact of the magnetic field) and
  inserted it into Eqs.~(\ref{two-photon-full}), (\ref{time-int}), and
  (\ref{Larmor}), respectively.} 
\end{figure}

Let us insert a set of parameters which are potentially realizable
with present or near-future technology \cite{design}.
Assuming an optical laser beam with a photon energy of 2.5 eV 
(in the laboratory frame) and a boost factor of 
$\gamma=300$, 
the photon energy in rest frame of the electrons 
$\omega=1.5\,\rm keV$ 
is still much smaller than the electron mass.  
With a laser intensity of order 
$10^{18}\,\rm W/cm^2$ 
in the laboratory frame, the electric field $E$ lies a factor of 
$1000$ 
below the Schwinger limit $E_S$ in the rest frame of the electrons 
and their transversal quivering motion 
$\f{\dot r}_{\rm cl}^2\approx1/9$ 
is still approximately nonrelativistic. 
After 100 laser cycles (half-width of Gaussian pulse), we obtain a
two-photon probability of order 
$10^{-11}$ 
frome one electron. 
Depending on their direction, the sum of the energies of the created 
photons in the laboratory frame is then around 
500~keV \footnote{The photons could be detected with Ge strip
  detectors (based on Compton scattering), which provide a good energy
  resolution (of order keV) in the range between 100 and 500 keV and
  are even sensitive to the polarization. A typical segmentation size
  of order millimeter results in an angular resolution of
  $\delta\vartheta=\ord(10^{-4})$ after a distance of order ten
  meters, which should be sufficient for a boost factor of
  $\gamma=300$, see Fig.~\ref{figure1}. 
}.  
The total probability for the competing classical counterpart 
(Larmor radiation) is much higher $\ord(10^{-2})$.
Fortunately, the monochromatic character (rest frame of electrons) 
of the Larmor radiation
(which just corresponds to Thomson scattering of the laser photons) 
in our set-up ensures that 
the phase-space regions of the two effects are very different.
In the laboratory frame, the phase space is quite distorted after the
boost, see Fig.~\ref{figure1}, but it is (at least in principle) still
possible to discriminate the two effects via suitable apertures and
energy filters etc. 

So far, we considered the case of single electrons only.
For many electrons, their space-time distribution and the resulting 
spatial interference becomes important (in addition to the temporal 
interference, which yields the resonance conditions
$k=\omega$ and $k+k'=\omega$, respectively).
For both, classical and quantum radiation, one should distinguish two 
major limiting cases: incoherent or coherent superposition.
If the electrons are randomly distributed and their typical distance 
is much larger than $1/\omega$, we have an incoherent superposition
(addition of probabilities).
For example, sending such a pulse of $N_e=10^9$ independent electrons
into a laser beam with the values discussed above, we obtain around
one Unruh event in hundred shots.  
Of course, a coherent superposition (constructive interference of 
amplitudes) would be much more effective.
One possibility to achieve the necessary phase coherence could be to 
confine all the electrons to within half a wavelength $\pi/\omega$. 
For optical lasers, this is probably hard to do. 
However, in an other system, an analogous spatial phase coherence has
been achieved already:
In undulators for free-electron lasers (FEL) based on 
self-amplification of spontaneous emission (SASE), the original 
electron pulse is split up into many nearly equidistant micro-bunches
via the back-reaction of the Larmor radiation.
These micro-bunches contain a significant fraction of the total number
of electrons and occur at distances equal to half the undulator period
(in the frame of the electrons).
Therefore, the amplitudes generated by the zig-zag motion of 
these micro-bunches interfere constructively in forward 
(i.e., electron beam) direction -- leading to the amplification of
Larmor radiation 
(in the ideal case $\propto N_e^2$ instead of $\propto N_e$). 
Comparing Eqs.~(\ref{time-int}) and (\ref{Larmor}), we see that the
quantum (two-photon) amplitudes (\ref{two-photon}) do also interfere
constructively if both photons (with $k+k'=\omega$) are emitted in 
forward direction. 

Let us estimate the order of magnitude for a realistic set of
parameters envisioned for near future facilities \cite{design}.
Sending a pulse containing $6\times10^9$ electrons with a boost factor
of $\gamma=4000$ into an undulator with a period of order ten
millimeters, the frequency in the electron frame is
$\omega=\ord(1\,\rm eV)$. 
For an undulator, the $K$-factor plays the role of the normalized
amplitude of the laser in Eq.~(\ref{normalized}) and is assumed to be 
below one.
After around 100 periods (undulator length of order one meter), we
obtain a single-electron Larmor probability (\ref{prob-Larmor})
on the percent level. 
However, the expected yield of $>10^{12}$ photons already indicates
that the $6\times10^9$ electrons do not radiate independently but
interfere constructively. 
Unfortunately, not all electrons of the pulse will behave coherently,
the effective fraction of electrons which are in phase is given by the 
bunching factor, which is also assumed to be on the percent level.  
Still, calculating the Unruh-Larmor ratio ${\mathfrak P}_{\rm Unruh}/
{\mathfrak P}_{\rm Larmor}^{(1)}=\ord(10^{-14})$, we would again
expect one Unruh pair in around one hundred shots \footnote{The
  monochromatic Larmor photons with an energy around 8~keV in the
  laboratory frame could be filtered out via multiple Bragg
  scattering. In order to eliminate further background, it might be
  useful to send the micro-bunches shaped in one undulator into a
  second one (and to get rid of the photons from the first undulator)
  and to switch on and off the undulator field smoothly, i.e., with a
  Gaussian instead of a rectangular envelope.}. 

By adding up the amplitudes generated by many electrons coherently, it 
might even be possible to reach the non-perturbative regime, where
multi-photon effects become important.
In this regime, there is a crucial difference between quantum and 
classical radiation:
Classical radiation can be described by a coherent state 
\bea
\ket{\alpha}=\exp\{\alpha\hat a^\dagger-\alpha^*\hat a\}\ket{0}
\,.
\ea
In this case, the photon number 
$\bra{\alpha}\hat n\ket{\alpha}=|\alpha|^2$
scales quadratically with the number of electrons $\alpha\propto N_e$ 
(constructive interference).
Quantum radiation, on the other hand, corresponds to a (multi-mode) 
squeezed state 
\bea
\ket{\xi}=\exp\left\{\xi\hat a_1^\dagger\hat a_2^\dagger
-\xi^*\hat a_1\hat a_2\right\}\ket{0}
\,,
\ea
where we consider two modes $\hat a_1$ and $\hat a_2$ for simplicity. 
For small amplitudes $\xi\ll1$, the photon number 
$\bra{\xi}\hat n_1\ket{\xi}=\sinh^2(|\xi|)$ also scales quadratically 
with the number of electrons $\xi\propto N_e$, but after a certain 
threshold $\xi=\ord(1)$ is reached, it grows 
exponentially \footnote{It is interesting to note that the
  single-photon distribution is thermal, i.e., the reduced density
  matrix $\hat\varrho_1$ of one photon obtained after averaging the
  above (entangled) quantum state over the other photon
  $\hat\varrho_1={\rm Tr}_2\{\ket{\xi}\bra{\xi}\}$ exactly corresponds
  to the canonical ensemble. The associated temperature, however, is
  not constant but depends on the quantum numbers of the photon.}.   
Ignoring all geometrical factors, the threshold can be estimated 
from Eq.~(\ref{two-photon}):  after passing 
\bea
N_e=\ord\left(\alpha_{\rm QED}^{-1}\frac{E_S}{E}\right)
\ea
electrons (in the oscillating micro-bunches), the two-photon
wave-packets start to grow exponentially 
(until their growth is limited by back-reaction etc.). 
Reaching this threshold is a quite ambitious goal, but may become
within reach with the next generation of free-electron lasers (FEL). 

The signatures of the Unruh effect discussed above bear strong
similarities to (spontaneous) parametric down-conversion 
\footnote{It should be mentioned here that x-ray down-conversion (in
  crystals, for example), which has already been observed in the
  laboratory \cite{x-ray}, could in principle also be interpreted as a
  signature of the Unruh effect (in the weak-field limit), since the
  involved electrons in the crystal can be considered
  quasi-free. However, the set-up discussed here offers more options
  for tuning (e.g., tapering of the undulator) and can be applied to a
  wider range of parameters. For instance, it involves much stronger
  fields (i.e., many laser photons interact coherently with the
  electrons) and one can reach higher energies (increasing $\gamma$
  and using an optical laser instead of an undulator etc.). Moreover,
  the efficiency is much larger, e.g., the considerations above show
  that it might even be possible to reach the (non-perturbative)
  regime where many photon pairs are created coherently (``two-photon 
  laser'').} 
known from quantum optics: 
The external periodic electromagnetic field corresponds to the pump
beam and the electrons are analogous to the nonlinear dielectric
medium. 
In both cases, the scattering properties (refractive index) of the
medium are varied periodically (frequency $\omega$) by the pump beam
and thereby the quantum vacuum fluctuations of the electromagnetic
field are converted into a pair of entangled photons (signal and
idler) whose energies add up to the pump frequency $k+k'=\omega$. 
In quantum optics, this mechanism is the main source for entangled
photon pairs which have a wide range of applications including 
concepts known from quantum information theory (e.g., tests of Bell's
inequality, quantum cryptography, or teleportation), 
two-photon interferometry,  
photonic Fock states (i.e., states with a well-defined photon number,
which could be used for counting excitations, for example),   
heralded photon emission, 
and coincidence experiments etc.
Since the quantum radiation discussed here consists of entangled
photon pairs 
\footnote{In case of spatial 
  interference, even the momentum of the photons could be controlled,
  i.e., they would be fully entangled in energy $(k+k'=\omega)$,
  momentum $\f{k}\|\f{k'}$, and polarization
  $(\f{e}_{\fk{k},\lambda}\cdot\f{e}_{\fk{k'},\lambda'})$. Thus, if
  one photon is detected, the quantum numbers of the other photon are
  fixed. 
} 
with much higher energies (which are more robust against some
disturbances and offer higher interaction rates), it may allow the
transfer of these quantum-optics applications into the multi-keV
regime (see also \cite{nuclear}). 

R.~S.~and G.~S.~were supported by the Emmy-Noether Programme of the
German Research Foundation (DFG, grant \# SCHU~1557/1-2).
The authors acknowledge support by the DFG cluster of excellence MAP 
and fruitful discussions with F.~Gr\"uner, U.~Schramm, J.~Schreiber,
and P.~Thirolf.    

$^*${\tt schuetz@theory.phy.tu-dresden.de}


\end{document}